\documentclass{article}
\usepackage{graphicx} % Required for inserting images
\usepackage{authblk}
\usepackage{amsmath}
\usepackage{multirow}
\usepackage[sort, comma, numbers]{natbib}
\usepackage{pdflscape}
\usepackage[skip=5pt plus1pt, indent=40pt]{parskip}

\usepackage{geometry}
\usepackage{fancyhdr}
\title{Quartered Chirp Spectral Envelope for Whispered vs Normal Speech Classification}
\author[1]{S. Johanan Joysingh}
\author[2]{P. Vijayalakshmi}
\author[3]{T. Nagarajan}
\affil[1]{Vellore Institute of Technology, Chennai}
\affil[2]{Sri Sivasubramaniya Nadar College of Engineering, Chennai}
\affil[3]{Shiv Nadar University Chennai}
\date{}

\begin{document}
\maketitle

\section{Abstract}
Whispered speech as an acceptable form of human-computer interaction is gaining traction.
Systems that address multiple modes of speech require a robust front-end speech classifier.
Performance of whispered vs normal speech classification drops in the presence of additive white Gaussian noise, since normal speech takes on some of the characteristics of whispered speech.
In this work, we propose a new feature named the quartered chirp spectral envelope, a combination of the chirp spectrum and the quartered spectral envelope, to classify whispered and normal speech.
The chirp spectrum can be fine-tuned to obtain customized features for a given task, and the quartered spectral envelope has been proven to work especially well for the current task.
The feature is trained on a one dimensional convolutional neural network, that captures the trends in the spectral envelope.
The proposed system performs better than the state of the art, in the presence of white noise.

\vspace{0.25cm}
\begin{center}
\textbf{Keywords}: \textit{whispered speech, classification, feature, human-computer interaction}
\end{center}

\section{Introduction}
Whispered speech is a common mode of speech in human-human interaction. 
It is used in all quiet scenarios such as at night, in the library, or on a secret law-enforcement mission, etc. 
But only in the last five years has it been considered seriously as a mode of human-computer interaction (HCI), after automatic speech recognition of normal speech matured. 
Besides the need of users to whisper to humans or machines, it is notable that a large community of people who have had a laryngectomy are only able to whisper.
In a report concluded in 2013, there were about 50,000 laryngectomees in the US alone \cite{brook2013laryngectomee}.
This means that addressing whispered speech in HCI is not only a luxury, but also a need of the society. 

\begin{figure}
    \centering
    \includegraphics[width=0.95\textwidth]{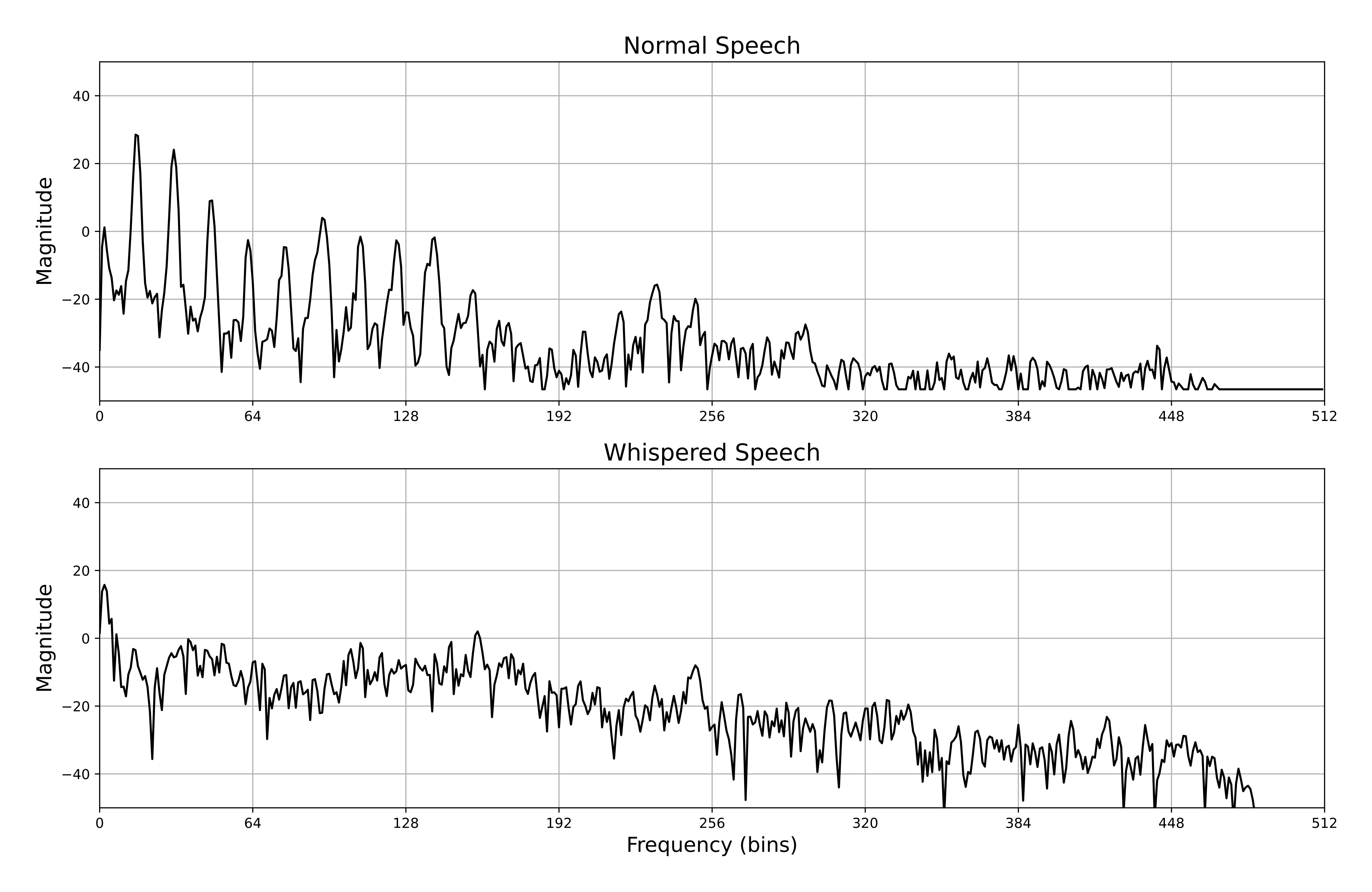}
    \caption{Difference between the spectral envelope of normal and whispered speech}
    \label{fig:spectrum-normalvswhisper}
\end{figure}

Whispered speech differs from normal speech mainly in the fact that it lacks pitch and pitch harmonics.
This is due to the lack of involvement of the glottis in whispered speech production. 
The log-magnitude spectrum of normal and whispered speech is illustrated in Figure \ref{fig:spectrum-normalvswhisper}.
The presence of pitch harmonics in normal speech, and its absence in whispered speech can be observed from this figure. 
Furthermore, this is especially evident in the first quarter, or the first 128 bins in Figure \ref{fig:spectrum-normalvswhisper}.

Whispered speech poses a specific problem in the presence of white noise.
Since the source of unvoiced sounds in a source-system speech production framework, is white noise, it is obvious how whispered speech can be very similar to white noise, in the absence of a role for the glottis.
A comprehensive analysis of whispered and normal speech can be found in \cite{wilson1985comparative} and \cite{wenndt2002study}, where a set of handcrafted features for their classification is also proposed.
A summary of these characteristics are that, in whispered speech: the locations of the formants are shifted higher, the amplitudes of the formants are lower, and the bandwidths of the formants are wider, when compared to normal speech.
% \begin{itemize}
%     \item the locations of the formants are shifted higher,
%     \item the amplitudes of the formants are lower, and
%     \item the bandwidths of the formants are wider,
% \end{itemize}
% when compared to normal speech.

Existing work on the classification of whispered speech include the use of features such as MFCC \cite{ashihara2019neural}, log filter-bank cepstral coefficients (LFCC), Teager energy cepstral coefficients (TECC) \cite{khoria2021teager}, log filter-bank energy (LFBE) \cite{raeesy2018lstm}, group delay spectrum \cite{shah2021exploiting}, and quartered spectral envelope (QSE) \cite{joysingh2023quartered}.
Classification is carried out using statistical and neural machine learning models such as, GMM \cite{zhang2009advancements}, LSTM \cite{raeesy2018lstm}, 2D-CNN \cite{ashihara2019neural, shah2021exploiting}, and 1D-CNN \cite{joysingh2023quartered}.

\section{Proposed Feature}
The feature proposed in the current work is the quartered chirp spectral envelope (QCSE).
It draws inspiration from our previous work in \cite{joysingh2023quartered}, which uses the quartered spectral envelope for whispered speech classification. 
The intuition behind the QSE feature in \cite{joysingh2023quartered} is that the absence and presence of pitch harmonics in whispered and normal speech respectively, is more evident in the first quarter of the spectral envelope, than the other three. 
The feature can be given as, 
\begin{equation}
    X(n,k) = S(n,k) \qquad 1 < k < K/4,
    \label{eq:qse}
\end{equation}
where $S(n,k)$ is the spectrogram with $n$ frames and $k$ frequency bins. 

\begin{figure}[ht]
    \centering
    \includegraphics[width=0.95\textwidth]{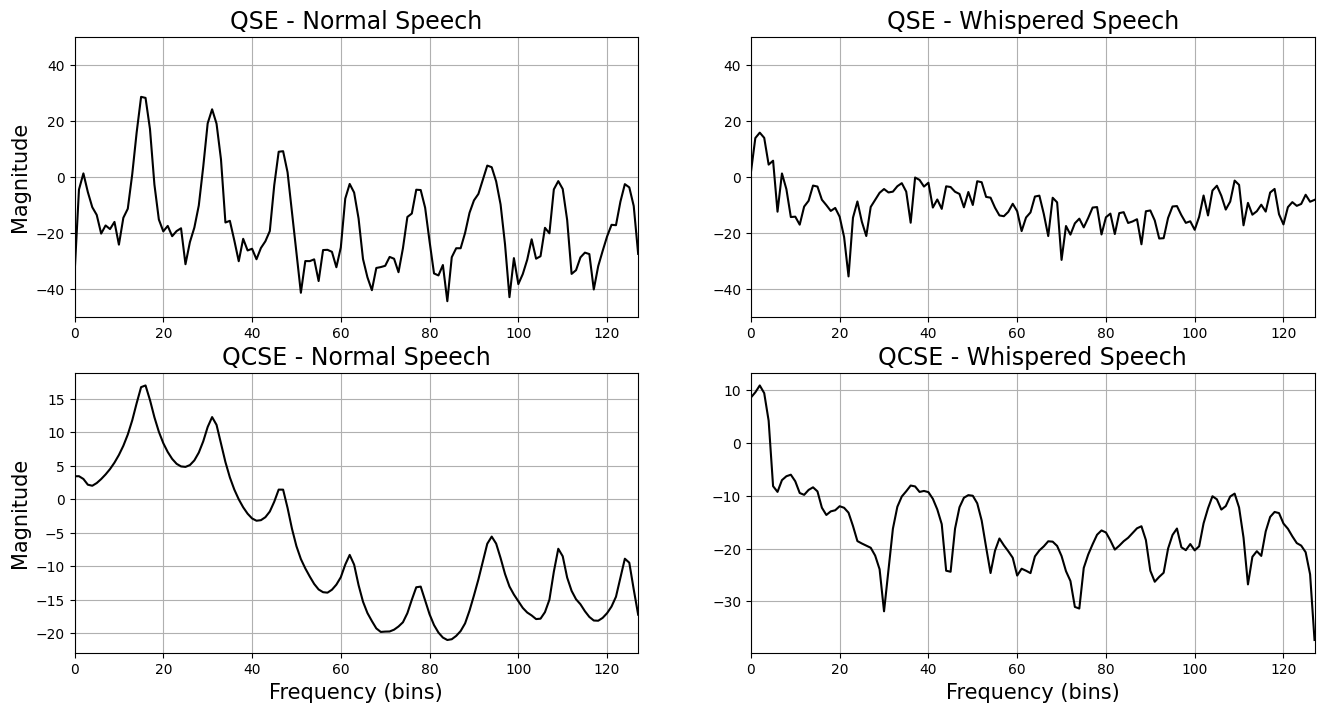}
    \caption{The QSE and QCSE (computed at $r=1.010$) of Normal and Whispered speech.}
    \label{fig:qse-qcse}
\end{figure}

\begin{figure}[ht]
    \centering
    \includegraphics[width=0.95\textwidth]{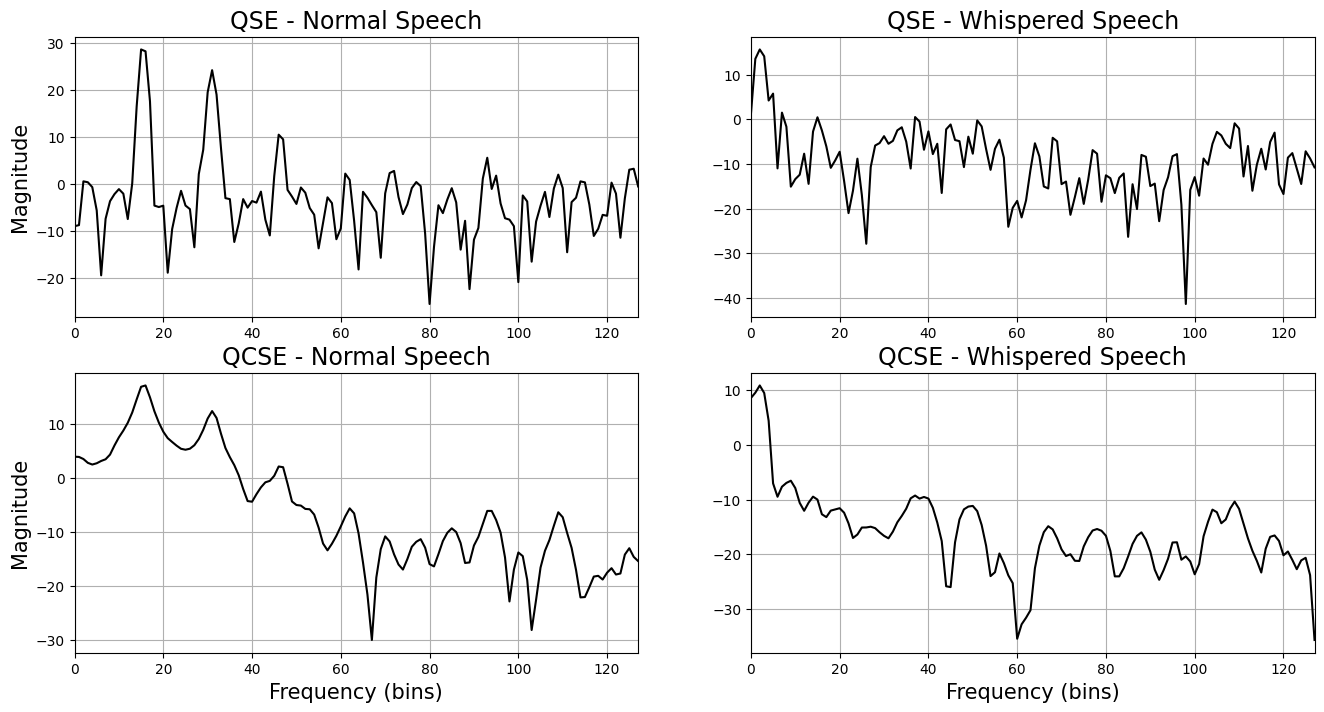}
    \caption{The QSE and QCSE (computed at $r=1.010$) of Normal and Whispered speech that is corrupted by noise at 5dB SNR.}
    \label{fig:qse-qcse-noise}
\end{figure}

% The chip spectrum as a feature for speech and audio applications was proposed in \cite{joysingh2024significance}. 
% It is based on the z-transform which is defined as, 
% \begin{equation}
%     X(z)|_{z=re^{j\omega}} = \sum_{n=0}^{N-1} x[n] (re^{j \omega})^{-n},
% \label{eq:ztransform_2}
% \end{equation}
% where $r$ is the radius of the circle on the z-plane, used for spectrum estimation. 
% The Fourier transform is the z-transform computed on the unit circle of the z-plane, that is, with $r = 1$ in Eq. \ref{eq:ztransform_2}.
% The chirp z-transform was proposed in \cite{rabiner1969chirp} which is the z-transform measured on any helical path on the z-plane, which includes circular paths. 
% The chirp spectrum as defined in \cite{joysingh2024significance} is the Fourier transform computed on any circle other than the unit circle in the z-plane, that is with radius $r \neq 1$.

The proposed feature also draws inspiration from our previous work on the chirp spectrum \cite{joysingh2024significance}.
The chirp spectrum in-turn is based on the chirp z-transform which was proposed in \cite{rabiner1969chirp}. 
The chirp z-transform is the z-transform measured on any helical path on the z-plane, which includes circular paths. 
The z-transform is defined as, 
\begin{equation}
    X(z)|_{z=re^{j\omega}} = \sum_{n=0}^{N-1} x[n] (re^{j \omega})^{-n},
\label{eq:ztransform_2}
\end{equation}
where $r$ is the radius of the circle on the z-plane, used for spectrum estimation. 
In the chirp z-transform, the radius $r$ is varied dynamically with respect to frequency, to trace helical or circular paths. 
The chirp spectrum as defined in \cite{joysingh2024significance} is the Fourier transform computed on any circle other than the unit circle in the z-plane, that is with radius $r \neq 1$ in Eq. \ref{eq:ztransform_2}.
It differs from the Fourier transform, which is the z-transform computed on the unit circle of the z-plane, that is, with $r = 1$.

% \begin{table*}[ht!]
% \begin{center}
% \caption{Details of the speech datasets used in the current work}
% \label{tab:dataset}
% \begin{tabular}{clcccc}
% \hline \hline
% \multicolumn{1}{l}{} & & \multicolumn{2}{c}{Train} & \multicolumn{2}{c}{Test} \\ \cline{3-6}
% \multicolumn{1}{l}{} & & \multicolumn{1}{l}{Normal} & Whisper & Normal & Whisper \\ \hline 
% \multirow{2}{*}{wTIMIT} & \multicolumn{1}{c}{Num. Utterances} & 10,934 & 10,245 & 725 & 723 \\
%  & Duration & \multicolumn{1}{l}{15h 31m} & 16h 25m & 1h 19m & 1h 28m \\ \hline
% \multirow{2}{*}{CHAINS} & \multicolumn{1}{c}{Num. Utterances} & 932 & 932 & 400 & 400 \\
%  & Duration & 1h 46m & 1h 48m & 13m & 13m \\ \hline \hline
% \end{tabular}
% \end{center}
% \end{table*}

\begin{table}[ht!]
\begin{center}
\caption{The number of utterances and the total duration of the two datasets considered in the current work.}
\label{tab:dataset}
\vspace{0.25cm}
\begin{tabular}{lcccc}
\hline \hline
 & \multicolumn{2}{c}{Train} & \multicolumn{2}{c}{Test} \\ \cline{2-5}
 & \multicolumn{1}{l}{Normal} & Whisper & Normal & Whisper \\ \hline 
\multicolumn{5}{l}{\textit{wTIMIT}} \\ \hline 
\multicolumn{1}{c}{Num. Utt.} & 10,934 & 10,245 & 725 & 723 \\
 Duration & \multicolumn{1}{l}{15h 31m} & 16h 25m & 1h 19m & 1h 28m \\ \hline
\multicolumn{5}{l}{\textit{CHAINS}} \\ \hline 
\multicolumn{1}{c}{Num. Utt.} & 932 & 932 & 400 & 400 \\
 Duration & 1h 46m & 1h 48m & 13m & 13m \\ \hline \hline
\end{tabular}
\end{center}
\end{table}

The intuition behind the chirp spectrum is that, since the strength of the spectral peaks caused due to singularities/sinusoidal sources are directly proportional to the proximity of the spectrum estimation \cite{rabiner1969chirp}, the radius $r$ can be pushed closer or further away from these singularities to obtain a sharper or smoother spectrum respectively, as required. 
In the case of speech that is adulterated by AWGN, the singularities of the signal are affected by those of the noise, which because of their lower strength lie further away from the unit circle and closer to the origin when compared to those of the signal.
If the spectrum is estimated slightly outside the unit circle, it will be influenced predominantly by the singularities of the signal rather than noise. 
% Since the singularities of the signal are present close to the unit circle but not on it, 
Hence, the chirp spectrum is computed at a radius $r = 1 + \Delta r$, where r = 0.01 in all the experiments that follow. 
% $1e^{-3} < \Delta r < 2e^{-2}$  
From the chirp spectrum, the chirp magnitude spectrum is derived, from which the QCSE can be determined using Eq. \ref{eq:qse} for the QSE.
It should be noted that the chirp spectrum can be computed using the fast Fourier transform (FFT), by rearranging Equation \ref{eq:ztransform_2} and writing it as,
\begin{equation}
    X(z)|_{z=re^{j\omega}} = \sum_{n=0}^{N-1} r^{-n}x[n] (e^{-j \omega n}),
\label{eq:ztransform_3}
\end{equation}
where the dot product of $r^{-n}$ and $x[n]$ is computed before computing the FFT.

\begin{figure*}[ht!]
    \centering
    \includegraphics[width=0.75\textwidth]{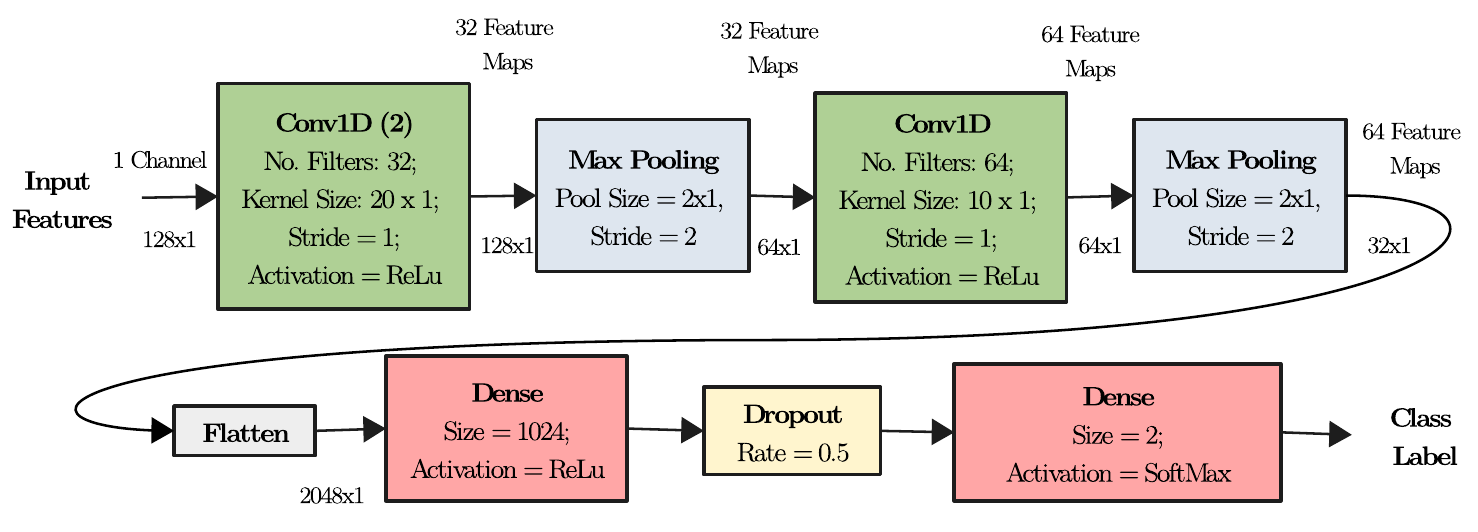}
    \caption{1D-CNN network architecture used in the current work}
    \label{fig:network}
\end{figure*}

\begin{table*}[h!]
\centering
\caption{Performance of all four systems considered for evaluation, at various signal-to-noise ratios, in terms of precision, recall and F1-measure of individual classes, and the overall accuracy, using the wTIMIT corpus.}
\label{tab:results}
\vspace{0.25cm}
\begin{tabular}{lrrrrrrrr}
\hline \hline
 & \multicolumn{1}{l}{} & \multicolumn{3}{c}{\textbf{Normal}} & \multicolumn{3}{c}{\textbf{Whisper}} & \multicolumn{1}{l}{} \\ \cline{3-9}
\textbf{Feature} & \multicolumn{1}{l}{\textbf{SNR (dB)}} & \multicolumn{1}{c}{\textbf{Pre}} & \multicolumn{1}{c}{\textbf{Re}} & \multicolumn{1}{c}{\textbf{F1}} & \multicolumn{1}{c}{\textbf{Pre}} & \multicolumn{1}{c}{\textbf{Re}} & \multicolumn{1}{c}{\textbf{F1}} & \multicolumn{1}{c}{\textbf{Acc}} \\ \hline
\multirow{3}{*}{QSE+1DCNN \cite{joysingh2023quartered}} 
 & 0 & 0.7872 & 1.0000 & 0.8809 & 1.0000 & 0.7289 & 0.8432 & 0.8646 \\ 
 & 5 & 0.8929 & 1.0000 & 0.9434 & 1.0000 & 0.8797 & 0.9360 & 0.9399 \\ 
 & 10 & 0.9474 & 0.9697 & 0.9584 & 0.9688 & 0.9461 & 0.9573 & 0.9579 \\ 
 &  & 0.9877 & 1.0000 & 0.9938 & 1.0000 & 0.9876 & 0.9937 & \textbf{0.9938} \\ \hline
\multirow{3}{*}{LFBE+LSTM \cite{raeesy2018lstm}} 
 & 0 & 0.9729 & 0.9393 & 0.9558 & 0.9412 & 0.9737 & 0.9572 & 0.9565 \\ 
 & 5 & 0.9673 & 0.9793 & 0.9733 & 0.9790 & 0.9668 & 0.9729 & 0.9731 \\ 
 & 10 & 0.9805 & 0.9710 & 0.9757 & 0.9712 & 0.9806 & 0.9759 & 0.9758 \\ 
 &  & 0.9783 & 0.9931 & 0.9856 & 0.9930 & 0.9779 & 0.9854 & 0.9855 \\ \hline
\multirow{3}{*}{QCSE+LSTM} 
 & 0 & 0.9516 & 0.9490 & 0.9503 & 0.9490 & 0.9516 & 0.9503 & 0.9503 \\ 
 & 5 & 0.9783 & 0.9310 & 0.9541 & 0.9340 & 0.9793 & 0.9561 & 0.9551 \\ 
 & 10 & 0.9831 & 0.9628 & 0.9728 & 0.9634 & 0.9834 & 0.9733 & 0.9731 \\ 
 &  & 0.9624 & 0.9876 & 0.9748 & 0.9872 & 0.9613 & 0.9741 & 0.9744 \\ \hline
\multirow{3}{*}{QCSE+1DCNN} 
 % & 0 & 0.9519 & 0.9559 & 0.9539 & 0.9556 & 0.9516 & 0.9536 & 0.9537 \\
 & 0 & 0.9803 & 0.9628 & 0.9715 & 0.9633 & 0.9806 & 0.9719 & \textbf{0.9717} \\
 % & 5 & 0.9694 & 0.9628 & 0.9661 & 0.9629 & 0.9696 & 0.9662 & 0.9662 \\
 & 5 & 0.9690 & 0.9903 & 0.9795 & 0.9901 & 0.9682 & 0.9790 & \textbf{0.9793} \\
 % & 10 & 0.9650 & 0.9890 & 0.9768 & 0.9887 & 0.9640 & 0.9762 & 0.9765 \\
 & 10 & 0.9863 & 0.9931 & 0.9897 & 0.9930 & 0.9862 & 0.9896 & \textbf{0.9896} \\
 &  & 0.9850 & 0.9972 & 0.9911 & 0.9972 & 0.9848 & 0.9910 & 0.9910 \\ \hline \hline
\end{tabular}
\end{table*}

The advantage offered by the chirp perspective, for the current task, is illustrated in Figure \ref{fig:qse-qcse} by comparing the QSE from \cite{joysingh2023quartered}, and QCSE from the current work.
% which shows the difference between the QSE and the QCSE (computed at $r=1.010$) for normal and whispered speech.
It can be seen from the figure that the difference between normal and whispered speech that is already evident in the QSE is more obvious in the QCSE.
It should be noted that the differences must be evaluated from the perspective of a 1D-CNN, which learns the contours in the spectral envelopes \cite{kiranyaz20211d}. 
In this context, the smoothed form of the QCSE of normal speech, when compared to the irregularities present in that of whispered speech, is proposed to give benefits in terms of classification accuracy.

Furthermore, the advantage offered by QCSE is more pronounced in the presence of AWGN.
% , as shown in Figure \ref{fig:qse-qcse-noise}. 
Figure \ref{fig:qse-qcse-noise}, shows the QSE and the QCSE (computed at $r=1.010$) of normal and whispered speech, both corrupted by AWGN at 5dB SNR.
It can be seen that the QSE of normal and whispered speech have started to look similar (again in the context of a 1D-CNN learning its contours). 
The QCSE on the other hand is able to maintain its overall smoothness considerably for normal speech, especially across the first 64 bins.
Furthermore, comparing Figures \ref{fig:qse-qcse} and \ref{fig:qse-qcse-noise}, it can be seen that for the QCSE of whispered speech, there is not much change with and without noise. 

\section{Experimental Setup and Results}
The details of the network architecture that is used to learn the proposed feature, and the datasets used for evaluation, are provided in the following sections. 

\subsection{Network Architecture}
The features are trained on a one dimensional convolutional neural network (1D-CNN) shown in Fig. \ref{fig:network}.
The proposed network needs to be computationally efficient, as a front-end classifier. 
It consists of two convolutional layers learning 32 and 64 filters. 
The kernel size of these filters are 20 and 10 for each of the two layers respectively.
It can be seen from Figures \ref{fig:qse-qcse} and \ref{fig:qse-qcse-noise} that they correspond approximately to the width of one major and one minor peak respectively.
The outputs of the convolutional layers are max pooled with a pool size of 2. 
% and dropped out with a ratio of 0.25.
The intermediate features produced by the 1D-CNN are flattened, and learnt by one hidden dense layer containing 1024 nodes. 
The number of parameters in this network is $\approx$2.9e6.

\subsection{Datasets}
The datasets used for evaluation are the wTIMIT \cite{lim2011computational} and the CHAINS (characterizing individual speakers) \cite{cummins2001chains} datasets.
The wTIMIT dataset is used in works such as \cite{khoria2021teager}, \cite{shah2021exploiting}, \cite{joysingh2023quartered}, and \cite{ashihara2019neural}, and the CHAINS dataset in \cite{khoria2021teager}, \cite{joysingh2023quartered}, and \cite{shah2021exploiting}.
The wTIMIT contains a large amount of parallel whispered and normal speech data. 
The CHAINS dataset, is meant for speaker identification, but since it contains a set of parallel normal and whispered speech data, it is repurposed for the current task.
The details of the datasets are provided in Table \ref{tab:dataset}.

\subsection{Evaluation}
During training the models are initialized with random weights and trained from scratch. 
They are trained until the loss stops improving. 
The checkpoints with the least two losses are used in testing, and the best out of these two are reported.
The evaluation is carried out at the utterance level. 
That is, each utterance is classified as either normal or whispered speech.
Frame-level scores are obtained from the network, and they are averaged to obtain the utterance level scores. 

The proposed system is compared with two state-of-the-art systems. 
The first uses a long-short term memory (LSTM) neural network, trained on log-filterbank energy (LFBE) features as in \cite{raeesy2018lstm} and \cite{ashihara2019neural}.
The second uses the QSE feature trained on 1D-CNN as in \cite{joysingh2023quartered}. 
Further more, to show the influence of 1D-CNN, the proposed QCSE feature is trained on the LSTM network as well for comparison. 
Each of the four systems considered for evaluation are trained and tested on both the wTIMIT and CHAINS dataset, at three different signal-to-noise ratios, namely 0dB, 5dB, and 10dB, and also without noise (clean speech). 
The systems are scored using precision, recall, and F1 measure for each class, along with the overall accuracy across two classes.
% The results are discussed in the section that follows.

\subsection{Results and Discussion}
\label{sec:results}
The results of the evaluation for the wTIMIT dataset are tabulated in Table \ref{tab:results}.
All algorithms produced 100\% accuracy for the CHAINS dataset, hence they are not tabulated.
Observing the F1 scores for both the classes, it can be seen that none of the systems are biased towards a particular class, hence just the overall accuracy is a good enough measure to compare all the systems in Table \ref{tab:results}.

First, it can be seen from the results that the QCSE+1DCNN system performs better than the LFBE+LSTM system across all signal-to-noise ratios.
% The evaluation of the proposed QCSE+1DCNN system against two current state-of-the-art systems shows that the proposed performs better than both existing systems. 
% This is true across all three SNR levels.
The same is true between the QSE+1DCNN and QCSE+1DCNN systems as well, except that the QSE+1DCNN system performs marginally better when there is no noise involved (0.9938 vs 0.9910).  
% Secondly, from the observed variance in the accuracy across multiple experiments of the same system, it can be noted that the QCSE+1DCNN system is more stable than any other system when exposed to noise at various levels. 
Secondly, from the observed variance in the accuracy of a particular system across multiple experiments carried out at various SNRs, it can be noted that the QCSE+1DCNN system is more stable than any other system. 
Meaning that the drop in performance when the signal is exposed to various levels of noise is not abrupt.
This is true to a degree in the case of the LFBE+LSTM system as well. 
Thirdly, The QCSE+LSTM system does not make much of an impact in the scores, but it does perform better than the QSE+1DCNN system, and is comparable to the LFBE+LSTM system, in the presence of noise.

\section{Conclusion}
The quartered chirp spectral envelope feature is proposed in the current work for the classification of normal and whispered speech. 
% The proposed feature is trained using a one dimensional convolutional neural network.
On comparing the proposed system with two existing state-of-the-art systems, namely the LFBE+LSTM system and the QSE+1DCNN system, in the presence of AWGN at three SNRs, it is shown that the proposed QCSE+1DCNN system offers improvement in all cases.
The improvement in performance is attributed to the fact that the chirp spectrum can offer a smoothened representation of the spectrum, when computed at a radius $r = 1 + \Delta r$, despite the presence of AWGN.
It is also observed from the results that the 1DCNN is better at capturing the patterns from the QCSE feature than LSTM. 
% From all the discussions so far, it can be concluded that the QCSE feature offers specific advantages in the presence of white noise, while the 1DCNN feature is better at capturing patterns from the QCSE feature.

%\bibliographystyle{IEEEtran}
%\bibliography{mybib}

\end{document}